\documentclass[prl,amsmath,twocolumn]{revtex4}
\usepackage{graphicx}
\usepackage{color}
\usepackage{amsmath}
\usepackage[american]{babel}

\begin{document}

\def\K{{\bf{K}}}
\def\Q{{\bf{Q}}}
\def\Gbar{\bar{G}}
\def\tk{\tilde{\bf{k}}}
\def\k{{\bf{k}}}
\def\Kp{{\bf{K}}^{\prime}}
\def\tp{t^{\prime}}

\title{Suppression of $d$-wave superconductivity in the checkerboard Hubbard model}
\author{D. G. S. P. Doluweera$^{1,3}$, A. Macridin$^{1}$, T. A. Maier$^{2}$, 
  M. Jarrell$^{1}$, Th. Pruschke$^{3}$}
\address{
$^{1}$ University of Cincinnati, Cincinnati, Ohio, 45221, USA\\
$^{2}$ Oak Ridge National Laboratory, Oak Ridge, Tennessee, 37831, USA\\
$^{3}$ Institut f\"ur Theoretische Physik, Universit\"at G\"ottingen, Friedrich-Hund-Platz 1,37077G\"ottingen, Germany \\}
\date{\today}
\begin{abstract}
Using a dynamical cluster quantum Monte Carlo approximation we investigate the $d$-wave superconducting transition temperature $T_c$ in the doped 2D repulsive Hubbard model with a weak inhomogeneity. The inhomogeneity is introduced 
in the hoppings $\tp$ and $t$ in the form of a checkerboard  pattern where $t$ is the hopping within a $2\times2$ plaquette and $\tp$ is the hopping between the plaquettes. We find inhomogeneity suppresses $T_c$. The characteristic spin excitation energy and the strength of $d$-wave pairing interaction decrease with decreasing $T_c$ suggesting a strong correlation between these quantities.
\end{abstract}
\maketitle

{\em{Introduction.}}
The role of inhomogeneity in High Temperature Superconductors (HTS) is still an unsettled issue. Neutron scattering experiments reveal the presence of 1D charge and spin microscopic inhomogeneities (stripes) in under-doped HTS \cite{tranquada,mook,sharma,hinkov} above the superconducting transition temperature $T_c$. Scanning tunneling microscopic imaging provide evidence for the presence of two dimensional real space modulations or local density of states modulations called checkerboard patterns in Bi-2212  \cite{howald-CB,hoffman-CB,vershin-CB,McElory-CB} and Na-CCOC \cite{hanaguri-CB,Hirai} superconducting samples. Inhomogeneities observed in cuprates led to theoretical scenarios for an inhomogeneity based pairing mechanism in HTS \cite{carlson,earrigoni-kivilson}. Other theoretical studies\cite{Ivar,karan,Tsai-Kivelson,loh-xy} argue that inhomogeneities can enhance pairing and $T_c$.

\begin{figure}[t]
\begin{center}
\includegraphics*[width=0.80in]{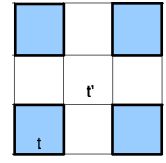}
\caption{(color on-line) $2\times 2$ Plaquette model. Here $t$ is intra-cluster hopping and $\tp$ is inter-cluster hopping. }
\label{fig:model}
\end{center}
\end{figure}

The  checkerboard  model\cite{Tsai-Kivelson} we investigate, is sketched in the upper panel of Fig.~\ref{fig:model}. It is homogeneous when $\tp= t$ and inhomogeneous when $\tp\neq t$. Here $t (\tp)$ is the hopping amplitude within (between) the cluster(s). The low-temperature phase diagram of the checkerboard Hubbard model (CBHM) in the strongly inhomogeneous limit $(\tp\ll t)$, i.e with weakly coupled clusters, was studied in Refs. \cite{Tsai-Kivelson,checker-nphase}. For  small $\tp\ll t$ and for values of on-site interaction $U\approx4.58t$, the authors find a $d$-wave superconducting state in addition to a variety of other phases. Their results led them to expect a maximum of $T_c$ at an optimal $\tp< t$. Other authors \cite{DCA_Tc,plqt-DMFT} find  $d$-wave superconducting state with a significant  $T_c$ for the homogeneous case $(\tp=t)$ in the 2D Hubbard   model. Therefore, it is interesting to see if the inhomogeneous CBHM has a higher $T_c$  compared to the homogeneous model. This is especially important in the weak inhomogeneity regime not addressed before, since experimentally observed inhomogeneities in the cuprates are weak.

In this paper we investigate $d$-wave superconductivity in the CBHM near the homogeneous limit, i.e.,  for  a weakly inhomogeneous system. We use the dynamical cluster approximation (DCA) \cite{DCA} to calculate $T_c$  as a function of $\tp$. DCA is a momentum($k$) space formulation and suitable to study the problem since our interest is around the homogeneous limit. We find that  a weak inhomogeneity suppresses $T_c$ . At  fixed values of $U/W$ and $U/t$ with doping $\delta$ appropriate for HTS, $T_{c}$  decreases for  $\tp \neq t$. Therefore, we find  a maximum $T_c$ in the homogeneous system. Furthermore calculations with fixed $U/t$ for $\tp<t$, show that $T_c$ decreases  proportional to the characteristic spin excitation energy and the strength of the $d$-wave pairing interaction, suggesting a strong correlation between them.

{\em{Formalism}.}
The DCA maps the lattice problem onto a periodic cluster embedded in a self consistent host. Short range correlations up to the linear cluster size $L_{c}$ are treated explicitly while longer range correlations are treated in a dynamical mean-field manner. We use a Quantum Monte Carlo (QMC) method to solve the cluster problem. We use the Maximum Entropy method to analytically continue imaginary time QMC data to real frequencies \cite{maxent}.

The Hamiltonian of our  model is
\begin{eqnarray}
H=- \sum_{\langle ij\rangle\sigma}t_{ij}\ \left(c^\dagger_{i\sigma} c_{j\sigma} +
c^\dagger_{j\sigma} c_{i\sigma}\right) + U \sum_i n_{i\sigma} n_{i -\sigma}\,.
\label{eq:h}
\end{eqnarray}
Where $ c^\dagger_{i\sigma} (c_{i\sigma})$ creates (destroys) an electron at site $i$ with spin $\sigma$ and $ n_{i\sigma}=c^\dagger_{i\sigma} c_{i\sigma}$. Here ${\langle ij \rangle}$ denotes the nearest neighbor sites $i$ and $j$. As shown in Fig. \ref{fig:model}, $t_{ij}=t (t_{ij}=\tp)$, when  $i$ and $j$ belong to the same (neighboring) plaquette(s). $U$ is the local Coulomb repulsion and it is site independent.

Here we present results obtained for a  $N_c=4$ sites cluster ($2\times 2$ sites).  $N_c=4$ is the smallest cluster which contains one $d$-wave plaquette and allows for a $d$-wave superconducting state \cite{DCA_Tc}. It has been intensively studied  in the last few years \cite{TM_review}. It has the advantage that the QMC sign problem is mild thus allowing the investigation of physics at low temperatures. However, because it contains only one $d$-wave plaquette it does not capture $d$-wave phase fluctuations and hence overestimates $T_c$\cite{DCA_Tc}. Calculations for larger clusters that capture these fluctuations will be an interesting future work. Details of the full DCA formalism applied to the CBHM  will be published elsewhere \cite{TbP}.

$T_c$ and the N\'eel temperature ($T_N$) are the temperatures where the superconducting and
antiferromagnetic susceptibilities diverge. For small clusters  it may be approximately 
described as the temperature at which the superconducting correlation length 
exceeds the linear cluster size. In a mean field calculation like DCA the transition temperature is 
overestimated  and decreases with increasing cluster size \cite{TM_review}. 
As we are treating the model in an elaborate mean-field scheme, the phase transition signalled by the divergence will quite likely not be present in the real 2D system, but may survive as Kosterlitz-Thouless transition, at least for d-wave superconductivity\cite{DCA_Tc}.

 $T_c$   can be obtained from the temperature dependence of the pairing matrix 
$M(K,K')=\Gamma^{pp}(K,K')\chi^{0}(K',K)$. Here, $K=(\K,{\bf{i}}\omega_n)$, $\Gamma^{pp}(K,K')$ (pairing interaction) is the irreducible particle-particle vertex and the bare bubble, $\chi^{0}(K',K)$, is the coarse-grained product of two fully dressed single particle Green's functions $G(k,k')$ and $G(-k,-k')$. $\omega_n=(2n+1)\pi T$ is the Matsubara frequency at temperature $T$. We determine $T_c$ from the temperature dependence of the leading eigenvalue $\lambda(T)$ of $M$ \cite{bulut,maier:pairprl}. We find $T_c$ from  $\lambda(T_c)=1$ and the symmetry of the superconducting state is determined by the symmetry of the corresponding  eigenvector, $\Phi(K)$, of $M$. For the investigated range of $\tp$, we find that the leading  eigenvector $\Phi_d(K)$ has the $d$-wave symmetry.

We define the strength of the pairing interaction $V_d$, by projecting out the pairing vertex $\Gamma^{pp}$, to the $d$-wave sub-pace as given in Eqn. \ref{eq:vdg}.
\begin{equation}
V_{d}(T) = \frac{1}{N_c} \sum_{\K\Kp} g(\K)\Gamma^{pp}(\K,\pi T;\Kp,\pi T)g(\Kp)
\label{eq:vdg}
\end{equation}
Here $g(\K)=\cos (\K_x)-\cos (\K_y)$ is the $d$-wave form factor. Correspondingly, we define the $d$-wave projected bare bubble
\begin{equation}
P_{d0}(T)=\frac{1}{N_c} \sum_{\K\Kp} g(\K)\chi^0_{pp}(\K,\Kp,\pi T) g(\Kp)
\label{eq:pd}
\end{equation}
$V_d$, $P_{d0}$ were initially introduced in Ref. \cite{maier:pairprb}. They satisfy the 
approximation  $V_d(T)P_{d0}(T)\approx \lambda_d(T)$.

{\em{Results}.}
Since the band width  $W=4(t+\tp)$ of the inhomogeneous system changes with $\tp$, 
we could use either $W$ or the maximum of $t$ and $t'$ as a unit of energy.  
Since both the Hamiltonian and $W$ are symmetric under the  interchange of  $t$ and $\tp$, 
in both cases the results should reflect this symmetry.
The first choice compensates for changes in $T_c$ due to changes of the kinetic energy.  
The second choice reflects the hopping suppression due to inhomogeneity
and will produce perfectly symmetric plots with abscissa  $\tp/t$ around one.

\begin{figure}[t]
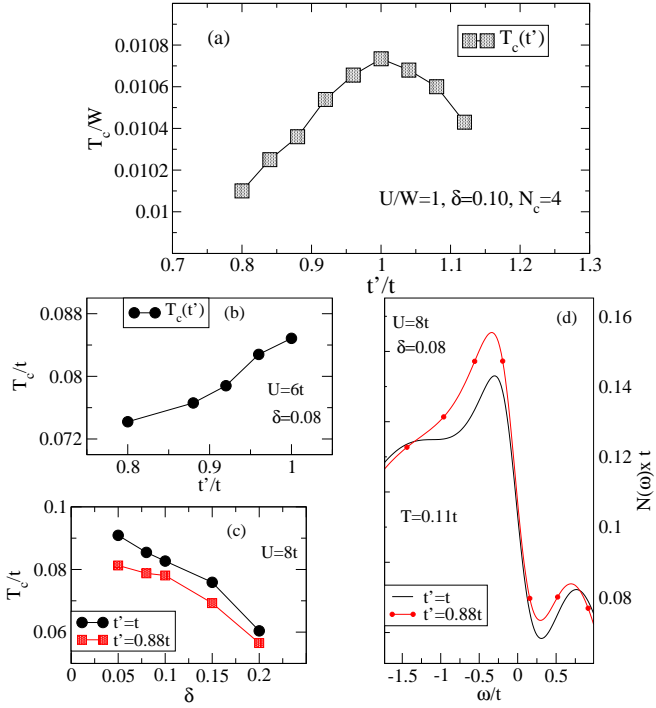

\begin{center}
\parbox[b]{2.55in}{\includegraphics*[width=2.7in]{./tcw.eps}}
\includegraphics*[width=3.4in]{./new-Tc-Dos.eps}
\caption{(color on-line) (a)$T_c(\tp)/W$ for $U/W=1$ and $10\%$ doping. $T_c$ is suppressed for $\tp\neq t$. (b) $T_c(\tp)/t$ for $U=6t$ and $8\%$ doping. $T_c$ monotonically decreases with decreasing $\tp$. (c) $T_c/t$ for $\tp=t$ and $\tp=0.88t$ as a function of doping $\delta$ when $U=8t$. (d) Low energy density of states $N(\omega)t$ for $\tp=t$ and $\tp=0.88t$ at $T=0.11t$ for $8\%$ doping and $U=8t$. Note the increase in low energy spectral weight for $\tp=0.88t$.}
\label{fig:lead}
\end{center}
\end{figure}

We find that  both of these choices lead to the same main conclusion that the inhomogeneity suppresses 
$T_c$. In Fig.~\ref{fig:lead} (a) we show results  when  $W$ is taken as the energy unit, for $U/W=1$ and at 10\% doping.
$T_c$ is maximum in the homogeneous limit $\tp/t=1$.
In the subsequent figures throughout the paper we will present results for the case where the energy unit is given by $t$.
As can be seen  from Fig.~\ref{fig:lead}(b), where  $U=6t$  at $8\%$ doping case is shown,
$T_c$ monotonically decreases with decreasing $\tp$.
The suppression of $T_c$ due to  inhomogeneity occurs on the doping range appropriate for HTS.  This can be seen in  Fig.~\ref{fig:lead}(c), where we show $T_c$ vs. $\delta$ \cite{Nc-4-Tcvsdel} for $U=8t$. Note that the inhomogeneous system with $\tp=0.88t$ has a lower $T_c$  than the the homogeneous system on the doping range $2\% -20\%$.

We find that $\tp \le t$ enhances the  density of states $N(\omega)$ at low excitation energy $(-0.5t<  \omega  <0.5t)$. 
This can be seen in Fig.~\ref{fig:lead}(d) where $N(\omega)$   for  $\tp=0.88t$ and $\tp=t$ are compared at  $8\%$ doping for $U=8t$. 
The  enhancement of  $N(\omega)$  is a band-structure effect, since it can also be noticed for the non-interacting problem,
i.e., when $U=0$.   The low energy part of $N(\omega)$ has an  important role in determining the value of $T_c$.

\begin{figure}[t]
\begin{center}
\includegraphics*[width=3.4in]{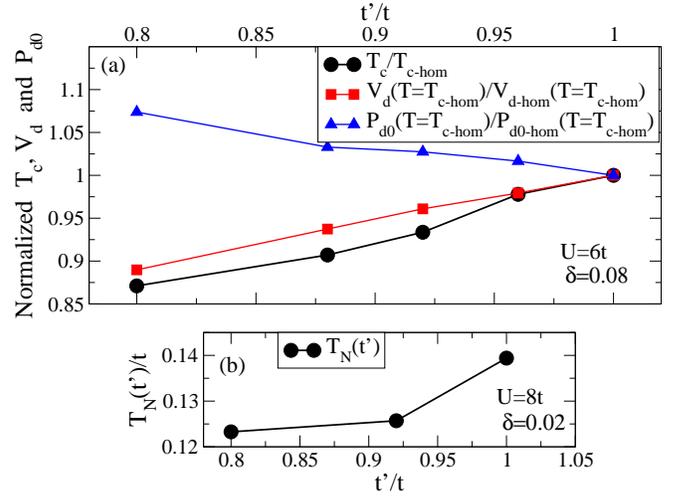}
\caption{(color on-line) (a) Normalized values of $T_{c}, V_{d}$ and $P_{d0}$ as a function of $\tp$ at  $8\%$ doping, $U=6t$ and $N_c=4$. $V_d$ and  $P_{d0}$ are described in the text. (b) N\'eel temperature $T_N/t$ for different values of $t^{\prime}$ at  $2\%$ doping for $U=8t$.}
\label{fig:TcVd}
\end{center}
\end{figure}

We find two competing effects of $\tp$ on the pairing matrix as captured by $V_d$ and $P_{d0}$. Fig.~\ref{fig:TcVd}(a) shows the 
 $\tp$ dependence of $V_d$, $P_{d0}$ and  $T_c$ for $U=6t$ at $8\%$ doping.  $V_d(\tp)$ and $P_{d0}(\tp)$ are shown at a temperature
 equal to the homogeneous $T_{c}$; ie. $T_{c-hom}$, and are normalized to their values at $\tp=t $ for easy comparison. 
 $V_d$  decreases with decreasing $\tp$, indicating a decreasing $d$-wave contribution to $\Gamma^{pp}$ as $\tp$ decreases. The effect of
 $\tp$ on  $\chi^0$ is captured by $P_{d0}$. $P_{d0}(\tp)$ increases with decreasing $\tp$. Since $P_{d0}$ is  related to $N(\omega)$
 through $G$, this behavior is related to the enhanced  low energy part of $N(\omega)$ discussed previously. 
The effect of $\tp$ on  $V_d$ and $P_{d0}$ provide a qualitative description of the behavior of $T_c$ since $V_d(T)P_{d0}(T)\approx \lambda_d(T)$. Note that  the relative decrease in $V_d$ is always larger than the relative increase in $P_{d0}$. Therefore, the net effect of $\tp$ is to reduce $T_c$.  The behavior of $V_d(\tp)$ and  $P_{d0}(\tp)$ at other values of $U$ and $\delta$ (not shown)
is similar, leading to a suppression of superconductivity.

Using the DCA for the homogeneous 2D Hubbard model, it was previously shown that the dominant part of the pairing interaction comes from the $S=1$ particle-hole magnetic channel \cite{maier:pairprl}. The energy scale of spin excitations is characterized by the super exchange interaction, $J=4 t^{2}/U$ and changing the hopping matrix elements of the homogeneous problem will also influence $J$. However, the relationship between $T_c$ and the characteristic energy scale of spin excitations, in the weakly inhomogeneous system is not known. Therefore it is worth studying the effect of $\tp$ on the N\'eel temperature and on the spin excitations as characterized by the magnetic structure factor $S(\Q,\omega)$

We find that at small doping,  $T_N$ decreases with decreasing $\tp$ as shown for $U=8t$  at $2\%$ hole doping in Fig.~\ref{fig:TcVd}(b). As discussed before, for a finite size cluster $T_N$ occurs when the antiferromagnetic correlation length $\zeta_{AF}$ exceeds the linear cluster size. Therefore decreasing $\zeta_{AF}$ indicates the suppression of antiferromagnetism in the weakly inhomogeneous system for small $\delta$ compared to a homogeneous system.

\begin{figure}[t]
\begin{center}
\includegraphics*[width=3.2in]{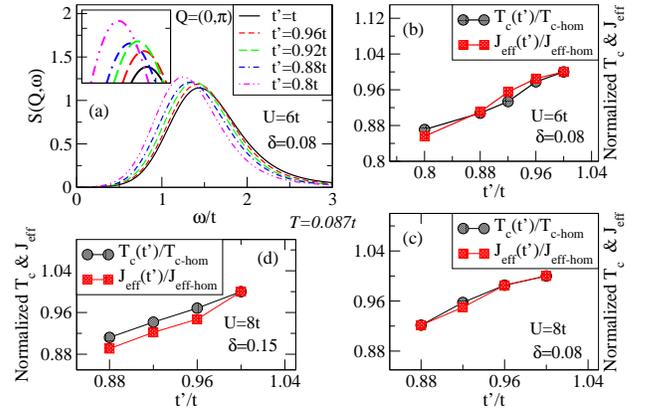}
\caption{(color on-line) (a) Magnetic structure factor $S(Q,\omega)$ for $U=6t, \delta=0.08$ and temperature $T= 0.087t$ at $Q=(0,\pi) $ for different values of  $\tp$. The location of the peak is a measure of the characteristic spin excitation energy, here defined as $J_{eff}$. Inset of (a) is a blow up to show the area around peak positions. $T_c(\tp)$ and  $J_{eff}(\tp,T=0.087t)$ normalized to their values at $\tp=t$, as a function of $\tp$ for (b) $U=6t, \delta=0.08$, (c) $U=8t, \delta=0.08$ and (d) $U=8t,\delta=0.15$. Note that both normalized $T_c(\tp)$  and  $J_{eff}(\tp)$ monotonically increase  with $\tp$.}
\label{fig:TcJ}
\end{center}
\end{figure}

The magnetic structure factor $S[\Q=(0,\pi),\omega]$ for various $\tp$ values is shown in Fig.~\ref{fig:TcJ}(a) for $U=6t$ at $8\%$ doping. The location of the peak of $S[\Q=(0,\pi),\omega]$  can be considered  a measure of the characteristic spin excitation energy
and we define its position to be $2 J_{eff}$. 
This is done  in analogy with linear spin wave theory\cite{j-explain, Manousakis} where the peak of the structure factor at $\Q=(0,\pi)$ is at energy $2J$, $J$ being  the  super exchange nearest neighbor spin interaction. 
Note that $J_{eff}(\tp)$ decreases with decreasing $\tp$, showing that a weak inhomogeneity soften the high energy spin excitations. This can be seen from the plot of  $S[\Q=(0,\pi),\omega]$ in Fig.~\ref{fig:TcJ}(a) and plots of $J_{eff}(\tp)$ in Figs.~\ref{fig:TcJ}(b), (c) and (d).

We find that $T_c\propto J_{eff}$, as suggested from Figs.~\ref{fig:TcJ}(b), (c) and (d).
By comparing $8\%$ doping Figs.~\ref{fig:TcJ}(b) and (c) with $15\%$ doping Fig.~\ref{fig:TcJ}(d), one can see that this proportionality holds better for small doping. The propotionality  between $T_c$ and the super exchange interaction $J$ are in accordance with the experimental observations reported in Ref. \cite{TcJns} and  the computational investigation on the homogeneous system in Ref. \cite{plqt-DMFT}.

{\em{Discussion}.}
The maximum $T_c$ occurs in the homogeneous system and it is suppressed by weak inhomogeneity. The suppression of $T_c$ prevails over the investigated range of doping $\delta$ and $U$, which are appropriate for a description of HTS. A recent experiment on Na doped Ca$_{2}$CuO$_{2}$Cl$_{2}$ (Na-CCOC) also indicates suppression of $T_c$ due to inhomogeneity over a wide range of doping \cite{Hirai}.

According to our analysis, the suppression of $T_c$ due to weak inhomogeneity is accompanied by a suppression of $V_d$ and $J_{eff}$. This is an important observation since it suggests a strong correlation between  the characteristic spin excitation energy $J_{eff}$ and the  pairing interaction. It is consistent with the previous finding that the pairing interaction in the 2D Hubbard model is dominated by the $S=1$ magnetic channel \cite{maier:pairprl} and other results reported by Maier {\em{et al.}}\cite{maier:pairprb,maier-spin-ss}. Therefore inhomogeneities that reduce magnetic contributions to the pairing interaction are likely to reduce $d$-wave superconductivity.

Since we find $T_c \propto J_{eff}$, it may be possible to increase
$T_c$ by enhancing $J_{eff}$. In our model we find that a weak inhomogeneity
decreases $J_{eff}$,  but other
kinds of inhomogeneity might increase $J_{eff}$ and thus, if our assumption is correct, enhance $T_c$ .  For example, weak on-site disorder seems to increase  $J$  in the vicinity of defects \cite{v-v enhancement}.

Regarding the speculation for the existence of an optimal inhomogeneity raised in Ref.~\cite{Tsai-Kivelson}, our results based on the $N_c=4$ cluster indicate that $T_c$ is an increasing function of $\tp$ in the weakly inhomogeneous system and the optimal $\tp$ occurs at $\tp=t$. Therefore our results for the weakly inhomogeneous system are not in agreement with this speculation , but they do not exclude the possibility for an optimal $\tp$ at smaller values of $\tp$. In addition, it will also be interesting to see how $T_c$ scales with cluster size $N_c$ when the weak inhomogeneity is present.

{\em{Conclusion}.}
Using the Dynamical Cluster Approximation for an $N_c=4$ cluster we find that a weak  inhomogeneity in the checkerboard Hubbard model suppresses $d$-wave superconductivity. The characteristic spin excitation energy  and the strength of the pairing interaction decrease along with decreasing $T_c$ suggesting a strong correlation between these quantities.

{\em{Acknowledgments}.}
We thank D. J. Scalapino for useful discussions. DGSPD, MJ and AM  acknowledge the grants NSF DMR-0706379 and DMR-0312680. AM acknowledges the grant CMSN DOE DE-FG02-04ER46129. The computation was carried out in the Ohio Supercomputer Center under project PES0611-1. DGSPD acknowledges the hospitality of the Institute of Theoretical Physics, where this work was started. TAM acknowledges the Center for Nanophase Materials Sciences, which is sponsored at Oak Ridge National Laboratory by the Division
of Scientific User Facilities, U.S. Department of Energy. TP acknowledges support by the German Science Foundation (DFG) through the collaborative research center SFB 602.


\begin{thebibliography}{99}
\bibitem{Nc-4-Tcvsdel} The $T_c$ vs. $\delta$ phase diagram of the homogeneous system found with $N_c=4$ cluster has been discussed in Ref. \cite{DCA-Nc4-Jarrell}.
\bibitem{j-explain} It is reasonable to view HTS as doped anti-ferromagnetic  Mott-insulators \cite{doped-mott} and magnetic excitations may be approximated by the magnon dispersion given by linear spin wave theory. These magnon excitations show a peak at $\Q=(0,\pi)$ at an energy $\omega = 2J$ \cite{Manousakis}.
\end{thebibliography}
\end{document}